\documentclass[12pt]{article} 
\usepackage{amssymb}
\usepackage{graphicx}
\begin{document} 
\begin{center}
{\large \bf The robust impact parameter profile of inelastic collisions}

\vspace{0.5cm}                   

{\bf I.M. Dremin}

\vspace{0.5cm}                       

{\it Lebedev Physical Institute, Russian Academy of Sciences, 
Moscow 119991, Russia
\footnote {Also at the National Research Nuclear University "MEPhI"}

e-mail: dremin@lpi.ru}\\

\end{center}

\begin{abstract}
It is shown that the impact parameter profile of inelastic hadron collisions
is robust to admissible variations of the shape of the diffraction cone of
elastic scattering. This conclusion is obtained using the unitarity condition
and experimental data only with no phenomenological model inputs.
\end{abstract}

The impact parameter profile of inelastic high energy hadron collisions is
determined as the probability of such reactions to take place at definite
impact parameters at a given energy (see, e.g., \cite{drZ}). It can be derived 
from the unitarity condition if the properties of the elastic scattering 
amplitude are known. We show that its general features are robust to variations 
of the shape of the differential cross section of elastic scattering with the 
transferred momentum and total energy measured experimentally.

The impact parameter profiles of elastic and inelastic hadron collisions are 
not directly measurable but they help us visualize the geometrical picture of 
partonic interactions indicating their space extension and the intensity. Our
intuitive guesses about the space-time development of these processes can be
corrected in this way.
The inelastic profile $G(s,b)$ is a function of the energy $s=4E^2$, where $E$ 
is the total energy of colliding particles in the center of mass system, and of 
the impact parameter $b$, which represents the transverse distance between their
centers. It is determined from the unitarity condition in a following way
\begin{equation}
G(s,b)=2{\rm Re}\Gamma (s,b)-\vert \Gamma (s,b)\vert ^2,
\label{unit}
\end{equation}
where
\begin{equation}                                                            
i\Gamma (s,b)=\frac {1}{2\sqrt {\pi }}\int _0^{\infty}d\vert t\vert f(s,t)
J_0(b\sqrt {\vert t\vert })
\label{gamm}
\end{equation}
is the elastic profile defined by the Fourier-Bessel transform  of the elastic 
scattering amplitude $f(s,t)$ which depends on energy and the transferred 
momentum squared
\begin{equation}
-t=2p^2(1-\cos \theta ) 
\label{trans}
\end{equation}
with $\theta $ denoting the scattering angle in the center of mass system
and $p$ the momentum.

The left-hand side of (\ref{unit}) called the overlap function describes the 
impact parameter profile of inelastic collisions of protons. Its widths shows
the spatial extension of the region of inelastic interactions. It satisfies the 
inequalities $0\leq G(s,b)\leq 1$ and determines how absoptive is the 
interaction region depending on the impact parameter (with $G=1$ for full 
absorption). If integrated over the impact parameters, (\ref{unit}) leads to
the general statement that the inelastic cross section equals to the difference
of the total and elastic cross sections.  

The differential cross section of elastic scattering $d\sigma /dt$ 
measured in experiments is 
related to the scattering amplitude $f(s,t)$ in a following way
\begin{equation}
\frac {d\sigma }{dt}=\vert f(s,t)\vert ^2.
\label{dsdt}
\end{equation}
The shape of $d\sigma /dt$ varies with energy. However, there are some common
features typical at high energies. Particles are elastically scattered mostly
at small transferred momenta within the so-called diffraction peak. It is
roughly approximated by the exponential shape
\begin{equation}
\frac {d\sigma }{dt}\propto e^{-B\vert t\vert }.
\label{expB}
\end{equation}
with the slope $B$ depending on energy $s$ and slightly varying with the 
transferred momentum $t$. 

Moreover, the real part of the amplitude is small compared to the imaginary
part within the diffraction cone at high energies. At the LHC, their ratio in 
forward direction $\rho _0$ is equal to 0.1 \cite{totem2}. It decreases within 
the cone and crosses the abscissa axis according to all phenomenological 
models and general statements of 
Ref. \cite{mar1}. That is why it is possible to neglect this ratio in Eq.
(\ref{gamm}) where it enters weighted by the suppressing exponential factor. 
The corresponding corrections to $G(s,b)$ are quadratic in $\rho $. Surely,
they are smaller than one percent and will not be considered in what follows.

If one neglects for some time by the dependence $B$ on $t$, 
the inelastic profile looks as
\begin{equation}
G(s,b)= \frac {2}{Z}e^{-\frac {b^2}{2B}}-\frac {1}{Z^2}e^{-\frac {b^2}{B}},
\label{ge}
\end{equation}
where $Z=4\pi B/\sigma _t$.

It is important that at any high energy from ISR to LHC the differential cross 
section becomes 4 or 5 orders of magnitude smaller before the exponential
regime (\ref{expB}) is replaced by another slower decreasing behavior at
larger transferred momenta (the Orear region). Therefore, the role of this tail
is negligible for the profile $G(s,b)$ since its contribution to the integral in
$\Gamma (s,b)$ (\ref{gamm}) is extremely small. 

The variations of the slope within the diffraction cone can be only important.
As was observed in experiments, they are twofold. The slope itself can change 
its value with the transferred momentum or/and there appear some oscillations 
imposed over its smooth shape.
At ISR energies, it was shown \cite{hol1, hol2, amal, carr} that the slope
becomes smaller at $\vert t\vert >0.12 - 0.15$ GeV$^2$ and the exponent in
(\ref{expB}) can be approximated more accurately by $Bt+Ct^2$ with positive $C$
or by the sum of two exponential terms with exponents differing by about
1.5 GeV$^{-2}$. The accuracy of the data is not enough to distinguish between
these fits. At LHC energies, the slope becomes larger at $\vert t\vert > 0.36$ 
GeV$^2$ \cite{totem1} so that $C<0$ or, in the case of two exponential terms, 
the exponents differ approximately by the same amount but with the opposite 
sign. Anyway, the impact of these variations on the inelastic 
profile at the LHC is very small as shown in Fig. 1a of Ref. \cite{ads} where 
its shapes are calculated either directly from experimental data or from their 
simple approximation by (\ref{expB}). They are almost indistinguishable.

Another interesting feature of the slope behavior was studied at the energy
$\sqrt s \approx 11$ GeV in Refs \cite{ant1, ant2} reviewed in Ref. \cite{zrt}.
Slight oscillations with $t$  in the behavior of $B$ at the level of 5 - 10 \%
were noticed. Some decline from the simple exponential form can be also seen
at ISR energies if carefully studied. It is intended to be studied with more
precision again at Protvino energies about 11 GeV. This effect should be looked 
for at the LHC energies as well. 

The corrections $\Delta G$ to the profile $G(s,b)$ are connected with the 
corrections $\Delta \Gamma $  to $\Gamma $ in a following way
\begin{equation}                        
\Delta G(s,b)=2\Delta \Gamma (1-\Gamma )=2\Delta \Gamma (1-\frac {1}{Z}\exp
(-b^2/2B)). 
\label{dgdg}
\end{equation}                                                          
At the LHC, where $Z=1$, no corrections appear at the center $b=0$ but all of 
them are shifted to the tail of the impact parameter distribution. That shows
their peripheral origin.

The impact of oscillations on the behavior of $G(s,b)$ can be estimated if we 
approximate this decline by the simplest oscillating function inserted in 
(\ref{gamm}), (\ref{unit}) so that $G(s,b)$ is changed by the amount
\begin{eqnarray}                        
\Delta G(s,b)= \nonumber \\
a(1-\frac {1}{Z}\exp (-b^2/2B))
\int _0^{\infty}d\vert t\vert \exp (Bt/2)J_0(b\sqrt {\vert t\vert })
\cos (\kappa (\vert t\vert -\vert t_0\vert)) = \nonumber \\
a(1-\frac {1}{Z}\exp (-b^2/2B))
\frac {2B}{B^2+4\kappa ^2}\exp (\frac {-b^2B}{2(b^2+4\kappa ^2)})[\cos u -
\frac {\kappa }{2B}\sin u], 
\label{del}
\end{eqnarray}                                                          
where $u=\kappa (\vert t_0\vert +\frac {b^2}{B^2+4\kappa ^2})$. The amplitude 
$a$, positions of zeros and period of oscillations are estimated from
approximations of Figures shown in Refs \cite{ant1, ant2, zrt}. They are
$a\approx 0.1;\; x_0^2\approx 0.07$ GeV$^2$; $\; \kappa \approx 5\pi$ GeV$^{-2}$.
These corrections to the inelastic profile at the LHC with $B\approx 20$ 
GeV$^{-2}$ are of the order of one percent or
even less. They can reveal themselves at very high impact parameters where the
profile values are small. The oscillations were ascribed in Ref \cite{stts} to
the inelastic diffraction processes possessing the peripheral origin.

Thus, the main structure of the inelastic profile in proton collisions remains
quite intact. Its general feature at LHC energies is the widely spread black
region at $b\leq 0.5$ fm which reveals itself in properties of jets produced
in very high multiplicity events \cite{ads}. Even though the corrections are
small, the fine structure of the profile should be further studied. It can
open ways to identification of various classes of inelastic processes with
different regions of impact parameters. More precise data about the 
substructure of the diffraction peak in $t$-variable are necessary to relate
them with inelastic processes of different kinds.

Let us stress once more that the unitarity condition in combination with 
experimental data about elastic scattering within the diffraction cone was
only used without any reference to QCD ideas or phenomenological models.

This short note is inspired by the discussion with M.G. Ryskin at the 
conference NSQCD2014 in Gatchina. This work was supported by the Russian
Foundation for Basic Research (project nos. 12-02-91504-CERN-a and
14-02-00099) and jointly by the Russian Academy of Sciences and CERN.

\end{document}